# The Turing Pattern Transition with the Growing Domain and Metabolic Rate Effects

Shin Nishihara [1*] and Toru Ohira [1]

[1] Graduate School of Mathematics, Nagoya University.

Furocho, Chikusaku, Nagoya, 464-8602, Japan.

*Corresponding author: shin.kurokawa.c8@math.nagoya-u.ac.jp, ORCID: 0009-0003-2351-0644

## Abstract

This study examines how patterns on mammal body surfaces change as they transition from juveniles to adults and with seasonal variations. Our previous research suggests that patterns formed in infancy may fade due to the growing domain effects, typically linked to the body's surface expanding as it grows, but this transition is influenced by various factors. Here, we focus on how heat, which can change with body growth, affects this process. Generally, smaller organisms lose heat more easily due to their higher surface area-to-volume ratio, while larger organisms retain heat more effectively. In fact, thermoregulation during infancy is a crucial factor directly influencing survival.

We propose a theoretical model that incorporates both growing domain and metabolic rate effects to explain the mechanisms behind these pattern transitions. The model suggests that Turing patterns formed during juvenile stages disperse in adulthood due to domain growth and changes in metabolic rates affecting reaction rates. Numerical analysis shows that when both growing domain and metabolic rate effects are considered, patterns disperse more rapidly than when only growing domain effects are accounted for.

Furthermore, we discuss potential correlations among pattern transitions, growth, and thermoregulation mechanisms, emphasizing the role of metabolic rates in maintaining body temperature. Our findings shed light on the intricate relationship between growth, thermoregulation, and pattern transitions in mammals, possibly useful for further research in this field.

## Statements and Declarations

The authors declare that they have no known competing financial interests or personal relationships that could have appeared to influence the work reported in this paper.

## 1. Introduction

If parents and children looked noticeably different, it could be important to understand why these differences occur and what purpose they serve. Exploring these differences might lead to new discoveries or insights. Take, for example, the Chinese softshell turtle (*Pelodiscus sinensis*): the patterns on its plastron formed during the embryonic and juvenile stages vanish as the turtle matures. This change is so drastic that comparing only the plastron might make them seem like entirely different creatures. It's been suggested that this pattern transition could be theoretically explained by the Growing Domain (GD) effects (typically linked to the body's surface expanding as it grows), where the plastron's domain expands [Nishihara and Ohira, 2024]. However, unlike the situation with pattern transitions on the less visually noticeable ventral side of reptiles, many mammals display easily identifiable and distinct patterns that change over time. For instance, species like wild boars (*Sus scrofa*), lions (*Panthera leo*), sika deer (*Cervus nippon*), and tapirs (*Tapirus indicus* and *terrestris*) exhibit clearly distinguishable patterns on their bodies that shift as they mature. Unlike the non-Turing patterns found on the plastron of *P. sinensis*, these mammals often display Turing patterns such as spots and/or stripes during their juvenile stages, which then fade away as they grow older. In the case of adult sika deer, there's a seasonal cycle where similar spotted patterns seen in juveniles reappear in summer and vanish in winter. While there are slight variations, such as between Malayan and South American tapirs, where the juvenile patterns disappear in adulthood, adult tapirs typically have distinct white and black areas on their bodies. Despite these subtle differences among species, the consistent phenomenon is that patterns formed during juvenility tend to vanish in adulthood across these mammals.

The aim of this study is to explore the possibility that patterns on the body surfaces of juvenile mammals arise as a "byproduct" of certain survival strategies. For instance, if these patterns don't arise as a specific byproduct but instead help in camouflage to avoid predators, then it's probable that the patterns on lion cubs may not persist into adulthood as their role changes from prey to predator. Conversely, tapirs may not have a reason to deliberately erase the camouflage patterns formed during their calf stages, as they remain prey even as adults (it's worth noting that the visibility of prey to predators is speculative [Filip Jaroš, 2012]). Essentially, if these patterns are formed for camouflage, scenarios like these arise where it's challenging to justify that purpose. By conceptualizing the patterns formed during the juvenile stages as Turing patterns and considering theoretical elements that may lead to the dispersion of these Turing patterns, this study suggests exploring their survival strategies that may imply the formation and dispersion of patterns as "byproducts." Similar to the theoretical proposal for the pattern transition on the plastron of the Chinese softshell turtle [Nishihara and Ohira, 2024], it is proposed that the GD effects may contribute to the pattern transition in these mammals due to the clear growth of the domain. However, it was also mentioned that there were cases where patterns do not disperse despite the effects [Nishihara and Ohira, 2024], indicating the possibility of other mechanisms inducing pattern transition.

Here, we focus on the fact that thermogenesis is crucial to the survival of mammalian newborns [Bienboire-Frosini et al., 2023]. Considering that juveniles are typically smaller and lose heat more readily compared to adults, it's conceivable that there could be some variation in the thermoregulatory mechanisms between juveniles and adults. These differences in thermoregulation might consequently lead to changes in metabolic rate, which could in turn trigger pattern transitions. For example, when it comes to endothermic thermogenesis in mammals, research has focused on thermoregulation mechanisms involving brown adipose tissue (BAT) and muscle nonshivering thermogenesis [Berg et al., 2006; Bienboire-Frosini, et al., 2023; Gaudry et al., 2017a; Gaudry et al., 2017b; Nowack et al., 2019; Ricquier, 2005]. Particularly interesting is the fact that the thermoregulation mechanism in wild boars evolves as they mature [Bienboire-Frosini et al., 2023]. The theoretical elements of the GD effects, coupled with changes in metabolic rates triggered by differences in body size, may work together to bring about changes in patterns on the body surfaces of these juvenile and adult mammals.

The present study proposes that the Turing pattern transition on the body surfaces of mammals can be theoretically induced by the GD and/or Metabolic Rate (MR) effects associated with the domain growth.

## 2. Modeling of Turing Pattern Transition Induced by GD and/or MR Effects

The previous study [Nishihara and Ohira, 2024] modeled the phenomenon where the black patterns on the plastron of P. sinensis are formed during the embryonic and juvenile stages and gradually disappear in the adult stage, based on the GD effects in the reaction-diffusion system. However, it is not pattern formation induced by Turing Instability, and those black patterns cannot be considered as Turing patterns. Hence, in this study, we propose a model for the phenomenon where Turing patterns formed based on Turing instability can undergo transition. For example, lions (*Panthera leo*) and wild boars (*Sus scrofa*) exhibit patterns resembling Turing patterns, such as spots and stripes, during their juvenile stage, which disappear as they mature (specific differences in pattern colors can be interpreted based on the assumption of whether morphogen acts as an activator or inhibitor [Bard, 1981]). Furthermore, even in sika deer (*Cervus nippon*), the spotted patterns on their body surface transition between summer and winter, even in the absence of the domain-growth condition (*i.e.*, the phenomenon of the pattern forming and dispersing is repeated in the fixed domain). In other words, it is suggested that the model we should explore includes additional effects besides the GD effects.

Thus, we first analyze the influence of the GD effects on Turing patterns caused by Turing instability, illustrating specific reaction terms (however, it should be noted that these reaction terms are not mandatory but merely one example). The following reactions for certain substances $U$ and $V$ ($U_P$, for example, refers to the precursor of the substance $U$, while $U_M$ refers to, for instance, a metabolite of $U$. The same applies to $V_P$ and $V_M$) with $N \in \mathbb{N}$:

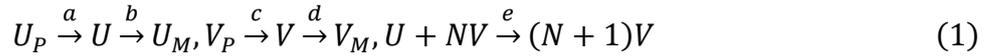

$$U_P \xrightarrow{a} U \xrightarrow{b} U_M, V_P \xrightarrow{c} V \xrightarrow{d} V_M, U + NV \xrightarrow{e} (N+1)V \qquad (1)$$

introduces the reaction terms below:

$$F(U,V) := -eUV^N + a - bU, G(U,V) := eUV^N + c - dV,$$

$$U[M], V[M], a\left[\frac{M}{\text{sec}}\right], b\left[\frac{1}{sec}\right], c\left[\frac{M}{sec}\right], d\left[\frac{1}{sec}\right], e\left[\frac{1}{M^N \cdot \text{sec}}\right]. \qquad (2)$$

With the reaction terms under the condition $N = 2$ (see Appendix-A for other values of $N$) and diffusion coefficients ($\Phi_U$ and $\Phi_V$ [$m^2/sec$]), a reaction-diffusion system on a spatially fixed domain is described as:

$$\frac{\partial U}{\partial t} = \Phi_U\left(\frac{\partial^2}{\partial X^2} + \alpha_u \frac{\partial^2}{\partial Y^2}\right)U + F(U,V),$$

$$\frac{\partial V}{\partial t} = \Phi_V\left(\frac{\partial^2}{\partial X^2} + \alpha_v \frac{\partial^2}{\partial Y^2}\right)V + G(U,V), \qquad (3)$$

where $\alpha_u, \alpha_v$ are constants that represent anisotropy (*i.e.*, $\alpha_u, \alpha_v = 1$ means isotropy), with the zero-flux boundary conditions applied. To nondimensionalize the system (3), the dimensionless quantities below are adopted: $u := (b/a)U$, $v := (b/a)V$, $t^* := t/k$, $\nabla^{*2} := (k\Phi_V/\phi_v)\nabla^2$, $\phi_u := (\Phi_U/\Phi_V)\phi_v$, $\beta := b^3/a^2e$, $\delta := b^2d/a^2e$, $\varepsilon := b^3c/a^3e$, $\mu := a^2ek/b^2$, which introduces the following nondimensionalized system:

$$\frac{\partial u}{\partial t} = \phi_u \nabla_u^2 u + f(u,v), \text{where } f(u,v) := \mu(-uv^2 - \beta u + \beta),$$

$$\frac{\partial v}{\partial t} = \phi_v \nabla_v^2 v + g(u,v), \text{where } g(u,v) := \mu(uv^2 - \delta v + \varepsilon), \qquad (4)$$

where the asterisks for $\nabla^2$ ( $\coloneqq \partial^2/\partial X^{*2} + \alpha.\partial^2/\partial Y^{*2}$ , where $X^*$ and $Y^*$ are the nondimensionalized spatial variables) and $t$ (*i.e.*, the nondimensionalized temporal variable) are omitted to simplify discussions, with the positive constants: $k[sec]$ and $\phi_v[1]$. When $\mu$ equals 1, the conditions for Turing Instability capable of forming Turing patterns are described as:

$$f_u^* + g_v^* < 0, \qquad f_u^* g_v^* - f_v^* g_u^* > 0,$$
$$\phi_v f_u^* + \phi_u g_v^* > 0, \qquad (\phi_v f_u^* + \phi_u g_v^*)^2 - 4\phi_u \phi_v (f_u^* g_v^* - f_v^* g_u^*) > 0, \qquad (5)$$

where $f_u = -v^2 - \beta, f_v = -2uv, g_u = v^2, g_v = 2uv - \delta$, while $f_u^*$, for instance, is defined as $\partial f/\partial u$ at the equilibrium point $(u^*, v^*)$ and the others are defined similarly (see Appendix-A for details). Hence, based on Turing Instability, the discrete wavenumbers $(k^2)$ that can form Turing patterns should exist in the following range of values $k_-^2$ and $k_+^2$ [Murray, 2013]:

$$k_\pm^2 \coloneqq \frac{1}{2\phi_u \phi_v} \Big( \phi_v f_u^* + \phi_u g_v^* \pm \sqrt{(\phi_v f_u^* + \phi_u g_v^*)^2 - 4\phi_u \phi_v (f_u^* g_v^* - f_v^* g_u^*)} \Big). \qquad (6)$$

Moreover, even when $\mu \neq 1$, the conditions described above are essentially unchanged, while the discrete wavenumbers' range is changed and described as $\mu k_\pm^2$. Based on the definitions of $\mu, \beta, \delta,$ and $\varepsilon$, it is technically possible to increase or decrease only $\mu$ in the system (4). For example, by reducing $b, d,$ and $e$ each by half, it is possible to double $\mu$ without altering $\beta, \delta, \varepsilon, k, \phi_u,$ and $\phi_v$. Increasing $\mu$ leads to an increase in wavenumber and a decrease in wavelength, while decreasing $\mu$ results in a decrease in wavenumber and an increase in wavelength. In other words, in this model, $\mu$ can be considered as the controller for adjusting the wavelength of Turing patterns, as mentioned in other studies [Murray, 2013; Woolley et al., 2021]. For instance, it is suggested that if you want to create large spotted patterns, you should set a small $\mu$, whereas if you want to form small spotted patterns, then you should set a large $\mu$. In this paper, $\mu$ can be understood as metabolic rates from the equations (1) and (4), where larger/smaller $\mu$ may suggest higher/lower metabolic rates.

We have discussed Turing Instability in a domain that does not grow with time, so far. However; actual organisms and tissues grow with time, so we construct a model by applying a domain that grows with time, the GD effects, to the model mentioned above. Whilst the study [Krause et al., 2019] reports that anisotropic growth demonstrates pattern evolutions, this study assumes isotropic growth. With the growing domain applied to the system (4) (*i.e.*, implying that the nondimensionalized spatial variables $X^*$ and $Y^*$ are time-dependent), the model with the GD effects is described as [Nishihara and Ohira, 2024]:

$$\frac{\partial u}{\partial t} = \frac{\phi_u}{r^2(t)}\left(\frac{\partial^2}{\partial x^2} + \alpha_u \frac{\partial^2}{\partial y^2}\right)u + f(u,v) - \frac{2r'(t)}{r(t)}u,$$

$$\frac{\partial v}{\partial t} = \frac{\phi_v}{r^2(t)}\left(\frac{\partial^2}{\partial x^2} + \alpha_v \frac{\partial^2}{\partial y^2}\right)v + g(u,v) - \frac{2r'(t)}{r(t)}v, \tag{7}$$

where $(X^*, Y^*) =: (r(t)x, r(t)y)$, $r(t)$ is a length-scaling function, and $r'(t) := \mathrm{d}r/\mathrm{d}t$. In order to simplify discussions, let $r(t)$ be defined as $\exp(d_a t/2)$, where $d_a$ is a positive constant, to ensure that the dilution terms do not become time-dependent [Krause et al., 2021] (this is merely an example for the sake of discussion simplification), and then it is assumed that the domain growth based on $r(t)$ starts from a certain time point $t_g$. Hence, during growing stages $(t \geq t_g)$ the system (7) is specifically described as:

$$\frac{\partial u}{\partial t} = \frac{\phi_u}{\exp\left(d_a(t - t_g)\right)}\left(\frac{\partial^2}{\partial x^2} + \alpha_u \frac{\partial^2}{\partial y^2}\right)u + \mu\left(-uv^2 - \left(\beta + \frac{d_a}{\mu}\right)u + \beta\right),$$

$$\frac{\partial v}{\partial t} = \frac{\phi_v}{\exp\left(d_a(t - t_g)\right)}\left(\frac{\partial^2}{\partial x^2} + \alpha_v \frac{\partial^2}{\partial y^2}\right)v + \mu\left(uv^2 - \left(\delta + \frac{d_a}{\mu}\right)v + \varepsilon\right). \tag{8}$$

Thus, if $d_a$ is selected such that the conditions for Turing Instability are met at $t = t_g$, then even as time progresses, those conditions will continue to be approximately satisfied as far as the linear theory allows. Additionally, the system (8) suggests that, as time elapses, larger wavenumbers (corresponding to smaller wavelengths) which can form Turing patterns start to be included in the range of $\mu k_{\pm}^2$. This case suggests the possibility that the Turing pattern formed until then $(t < t_g)$ does not disappear. Since the phenomenon discussed in this paper is the dispersion of formed patterns, we use and analyze $d_a$ from this point onwards, which no longer satisfies Turing Instability at $t = t_g$, suggesting the potential dispersion of Turing patterns.

However, the seasonal pattern transition on the body surface of sika deer is unlikely to be explained only by the GD effects. Furthermore, concerning the pattern transition in lions and wild boars, it would require unrealistic time frames for the patterns to disappear, potentially necessitating the unrealistic growth of the animals into giant adults. Therefore, we focus on the possibility of small $\mu$ inducing Turing patterns with very large wavelengths that could lead to pattern dispersion. Based on the principle of surface area-to-volume ratios, considering that juveniles may critically require more metabolic heat than adults [Bienboire-Frosini et al., 2023], the present study proposes that if it is assumed that the domain growth can decrease $\mu$, it would likely be when $\mu$ has a positive correlation with metabolic rates. Thus, in order to incorporate another equation for the temperature $T[\degree C]$ correlated with $\mu$ into system (8) following Pennes' Bioheat Equation, we consider the case where $Q(T)[W/m^2]$, the metabolic rate, is adjusted such that $T = T_{th}$, which is a certain constant, and since the perfusion term correlates with the metabolic rate, for convenience, it is defined herein as included within the logistic-type function, $Q(T)$, in this paper [Marn et al., 2019; Shrestha et al., 2020; Singh, 2024] (this study focuses on a two-dimensional domain [Othmer et al., 2009]), with the Dirichlet boundary condition applied: $T|_{\partial\Omega} = 0$. For example, the following equation including the perfusion term (*i.e.*, $\rho_b c_b w_b(T_{th} - T)$ in the equation (10) below) will be a possible consideration:

$$\rho c \frac{\partial T}{\partial(kt)} = \rho c \Phi_\tau \left( \frac{\partial^2}{\partial X^2} + \frac{\partial^2}{\partial Y^2} \right) T + K \rho c T_{th} \cdot T \left( 1 - \frac{T}{T_{th}} \right), \tag{9}$$

where $\rho[kg/m^2], c[J/(kg \cdot °C)], k[sec], \Phi_\tau[m^2/sec], K[1/(°C \cdot sec)]$ are positive constants, and although the same spatial variables are employed in this paper for the sake of simplification of discussions, it is important to recognize that such a constraint does not exist. Note that, as previously mentioned, the asterisk for $t$ has been omitted to simplify discussions and both the perfusion term and the metabolic rate term can be defined as indicated by the following equations, and then both terms are included in $Q(T)$ to express it as shown in the equation (9):

$$Q(T) = K \rho c T_{th} T \left( 1 - \frac{T}{T_{th}} \right) \coloneqq \rho_b c_b w_b(T_{th} - T) + K \rho c T_{th}' \cdot T' \left( 1 - \frac{T'}{T_{th}'} \right), \tag{10}$$

where $T' \coloneqq T - (\rho_b c_b w_b / K \rho c)$. Defining $\tau \coloneqq T/\lambda$ using a positive constant $\lambda[°C]$, $\kappa \coloneqq k K$ using a positive constant $\kappa[1/°C]$, $\phi_\tau \coloneqq (\Phi_\tau/\Phi_V)\phi_v$ using a positive constant $\phi_\tau[1]$, to nondimensionalize $T$ leads to the following equation:

$$\frac{\partial \tau}{\partial t} = \phi_\tau \nabla^2 \tau + q(\tau), \tag{11}$$

where $q(\tau) \coloneqq \kappa \lambda \tau(\tau_{th} - \tau)$ and $\tau_{th} \coloneqq T_{th}/\lambda, with \tau|_{\partial\Omega} = 0$, with the asterisk for $\nabla^2 \coloneqq \partial^2/\partial X^{*2} + \partial^2/\partial Y^{*2}$ omitted to simplify discussions, and we define the average $q(\tau)$, the metabolic rate, per unit area in the domain at each $t$ as $\bar{q}(\tau)$, and define $\hat{\mu}(\tau)$ for metabolic rate effects at the time as $\hat{\mu}(\tau) \coloneqq \gamma \bar{q}(\tau)$ where $\gamma$ is a positive constant as $\hat{\mu}(\tau)$ has a positive correlation with metabolic rates in the present study. The metabolic rate, $q(\tau)$, is defined under the assumption that metabolic heat is generated at all points, in the equation (11). Although there is no necessity to calculate the area-average value of $q(\tau)$, the metabolic heat generation on the boundary becomes relatively higher compared to the interior of the domain since the equation (11) imposes the Dirichlet boundary conditions. Note that, in this paper, the Dirichlet boundary conditions are imposed in such a way that significant changes in $\hat{\mu}(\tau)$ occur, in order to focus the discussion on the impact of $\hat{\mu}(\tau)$. This heterogeneity on the domain becomes a somewhat intricate point for discussing pattern

transitions, hence we use $\bar{q}(\tau)$ in this paper. Therefore, to explain the pattern transition on the body surfaces of the mammals mentioned earlier, the present study proposes the following model by incorporating the GD and MR effects:

$$\frac{\partial u}{\partial t} = \frac{\phi_u}{r^2(t)}\left(\frac{\partial^2}{\partial x^2} + \alpha_u \frac{\partial^2}{\partial y^2}\right)u + \hat{\mu}(\tau)\left(-uv^2 + \beta(1-u)\right) - \frac{2r'(t)}{r(t)}u,$$

$$\frac{\partial v}{\partial t} = \frac{\phi_v}{r^2(t)}\left(\frac{\partial^2}{\partial x^2} + \alpha_v \frac{\partial^2}{\partial y^2}\right)v + \hat{\mu}(\tau)(uv^2 - \delta v + \varepsilon) - \frac{2r'(t)}{r(t)}v,$$

$$\frac{\partial \tau}{\partial t} = \frac{\phi_\tau}{r^2(t)}\left(\frac{\partial^2}{\partial x^2} + \alpha_\tau \frac{\partial^2}{\partial y^2}\right)\tau + q(\tau) - \frac{2r'(t)}{r(t)}\tau,$$

$$\hat{\mu}(\tau) \coloneqq \gamma \bar{q}(\tau), \text{ where } \bar{q}(\tau) \coloneqq \frac{\int_\Omega q(\tau)dS}{|\Omega|} \text{ at each } t, \tag{12}$$

with the zero-flux boundary conditions applied for $u, v$, and $\tau|_{\partial\Omega} = 0$.

## 3. Result

We conducted numerical comparisons to observe how Turing patterns (spots) form in different systems. Specifically, we examined how patterns developed in the system (4) without the GD or MR effects compare to those in the systems (7) and (8) with only the GD effects, and in the system (12) with both GD and MR effects. Our numerical analysis reveals that even in the systems (7) and (8) with only GD effects, patterns disperse. However, in the system (12) with both GD and MR effects, the dispersion occurs more rapidly. In the latter case, the patterns disperse with a growth of 7.4 times, where in the former case the patterns are still being formed, while in the former case, the patterns disperse when the domain grows by approximately 16.4 times.

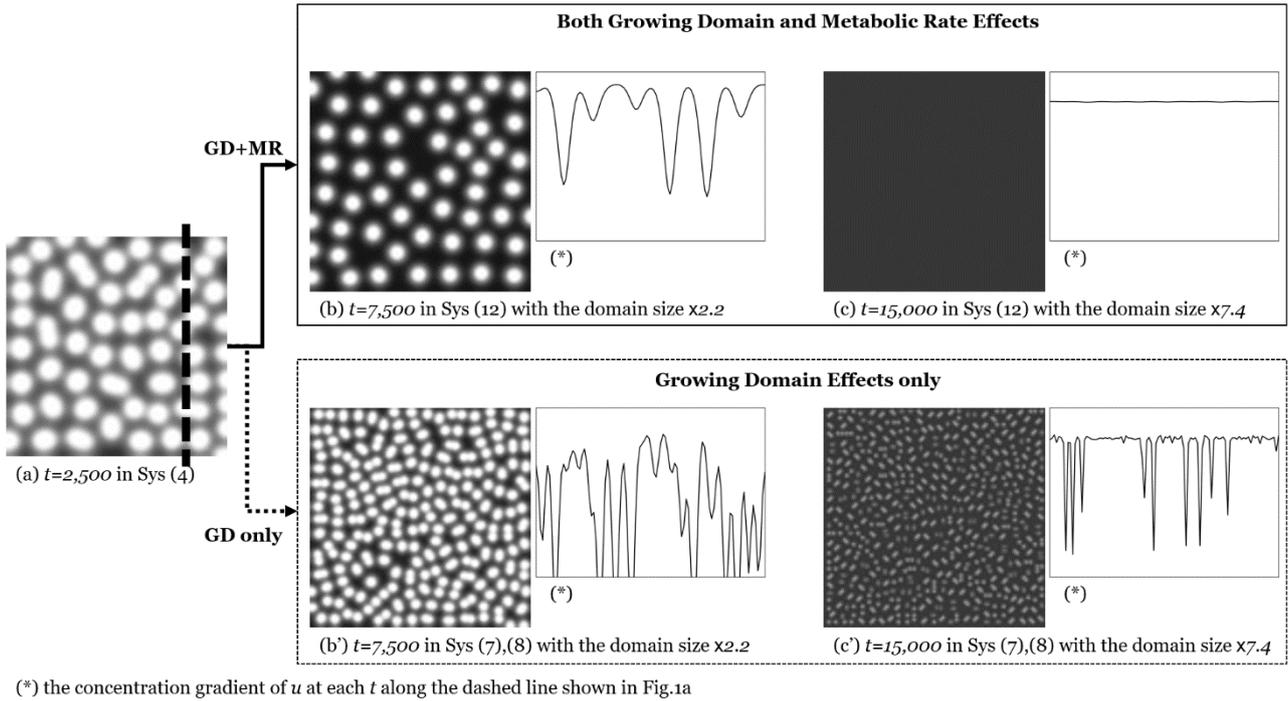

(*) the concentration gradient of $u$ at each $t$ along the dashed line shown in Fig.1a

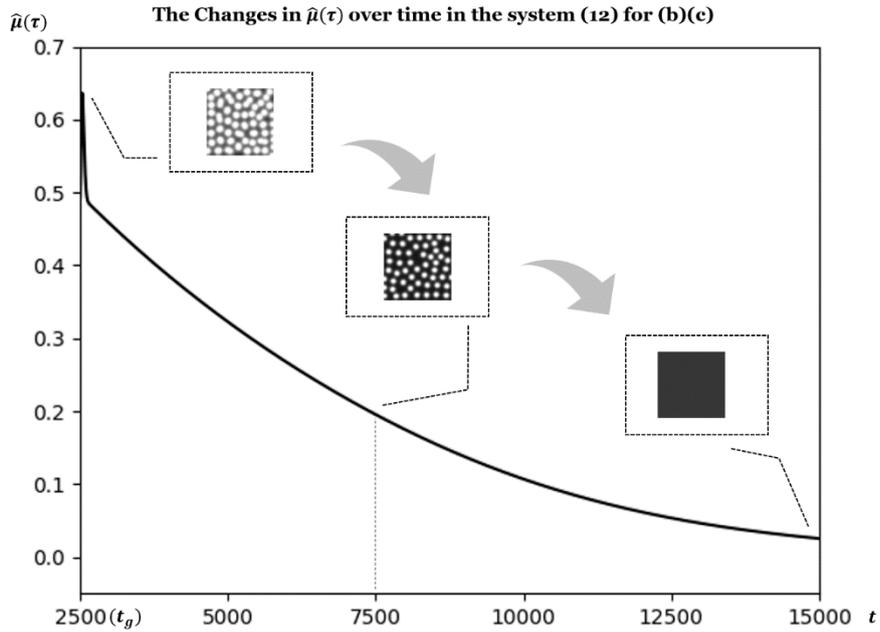

**Figure 1.** The numerical analysis results: (**a**) the pattern formed at $t = t_g = 2,500$ in the system (4), (**b**)(**c**) those at $t = 7,500$, and $15,000$ in the system (12) with the decrease in $\hat{\mu}(\tau)$, (**b'**)(**c'**) those at $t = 7,500$, and $15,000$ in the systems (7) and (8), under the parameters:

$(\phi_u, \phi_v, \beta, \delta, \varepsilon, d_a, \mu(\text{the initial value in the system (4)}), \alpha_u, \alpha_v, \phi_\tau, \alpha_\tau, \kappa\lambda, \tau_{th}, \gamma) =$
$(0.21, 0.05, 0.07, 0.18, 0.01, 3.2 \times 10^{-4}, 0.5, 1.0, 1.0, 0.24, 1.0, 0.2, 0.5, 300)$.

In addition, even in cases of slower growth (for instance, when $d_a = 2.0 \times 10^{-4}$), although requiring more time ($t = 24{,}500$) the patterns similarly disappear in the system (12), whereas in the systems (7) and (8), they are still formed at the same time point  (see Appendix-B).

As mentioned earlier, in the case of sika deer, their spotted patterns transition with seasonal changes without the growth of the domain. In other words, the pattern transition on sika deer's body surface may follow the system (13) described below, which removes the GD effect from the system (12). Note that the system (13) differs from the system (12) in that while the system (12) achieves pattern transitions without altering the coefficient values, the system (13) requires intentional modification of the coefficient $\kappa\lambda$, as will be discussed later. This necessity to intentionally change the value of coefficient $\kappa\lambda$ distinguishes the numerical analysis results obtained from the system (12) and those from the system (13).

$$\frac{\partial u}{\partial t} = \phi_u \left( \frac{\partial^2}{\partial x^2} + \alpha_u \frac{\partial^2}{\partial y^2} \right) u + \hat{\mu}(\tau)\left( -uv^2 + \beta(1-u) \right),$$

$$\frac{\partial v}{\partial t} = \phi_v \left( \frac{\partial^2}{\partial x^2} + \alpha_v \frac{\partial^2}{\partial y^2} \right) v + \hat{\mu}(\tau)(uv^2 - \delta v + \varepsilon),$$

$$\frac{\partial \tau}{\partial t} = \phi_\tau \left( \frac{\partial^2}{\partial x^2} + \alpha_\tau \frac{\partial^2}{\partial y^2} \right) \tau + q(\tau). \tag{13}$$

In this case, the MR effects (changes) are not induced as there is no GD effect. Hence, assuming that the efficiency of metabolic heat generation in sika deer decreases in the winter, decreasing the coefficient, $\kappa\lambda$, in $q(\tau)$ resulted in numerical analysis results showing the dispersion of the spotted patterns as follows.

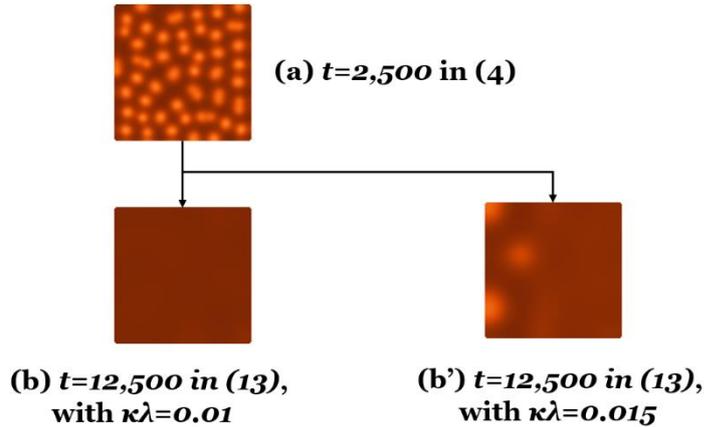

**(a) $t=2{,}500$ in (4)**

**(b) $t=12{,}500$ in (13),
with $\kappa\lambda=0.01$**          **(b') $t=12{,}500$ in (13),
with $\kappa\lambda=0.015$**

**Figure 2.** The numerical analysis results: (**a**) the pattern formed at  $t = t_g = 2{,}500$  in the system (4) under the same parameters as those of Fig. 1a, (**b**)(**b'**) those at  $t = 12{,}500$  in the system (13) with $\kappa\lambda = 0.01$, and  $0.015$, respectively. (Figures in Color)

## 4. Conclusion

In mammals, there are species that alter the patterns on their body surfaces during the process of transitioning from juveniles to adults, as well as with seasonal changes. In this paper, we theoretically and abstractly conceptualize this phenomenon, proposing a model in which Turing patterns initially formed transition due to the growth of the domain and the associated changes in metabolic rates. While the present study has demonstrated that Turing patterns can disperse due to the GD effects (dilution and diffusion coefficient decrease) as shown in the previous study [Nishihara and Ohira, 2024], comparing these results with actual phenomena has revealed unrealistic aspects (for example, requiring growth to a giant domain for pattern dispersion). Therefore, we focus on the possibility that metabolic rates relatively decrease within the grown domain to be able to efficiently maintain temperature within the domain due to growth, and propose a model incorporating the MR effects (decrease in metabolic rates). As a result, the dispersion of the patterns is confirmed in domains of realistic sizes more quickly than in cases with only the GD effect, due to the combined action of both the GD and MR effects. Furthermore, by applying only the MR effect for cases where the domain does not grow, *i.e.*, patterns transition with seasonal changes without the GD effect, pattern dispersion is reproduced by modifying the coefficient associated with the MR effect. In conclusion, the present study theoretically proposes that Turing patterns formed based on a certain chemical reaction can transition and disperse due to the growth of the domain (the GD effect) and changes in metabolic rates affecting reaction rates (the MR effect).

## 5. Discussion

First and foremost, it should be emphasized that the model proposed in this paper, the system (12), is not derived from solid experiments but rather is a theoretically proposed model that abstractly captures the transition of patterns on the body surfaces of mammals. It is undoubtedly agreeable that mammals undergo significant growth from juveniles to adults, and temperature is crucial for them. The former is self-evident, and regarding the latter, especially for newborns, thermoregulation is literally a matter of life and death [Bienboire-Frosini et al., 2023], and resolving this issue for survival is inevitable. For instance, in the case of wild boars, it has been confirmed that patterns transition during the growth process from juveniles to adults and that the mechanisms of thermoregulation change [Bienboire-Frosini et al., 2023]. While it's too early to directly connect these two aspects, it's possible that there could be some correlation, which would be a subject for future research. For example, in cold environments, prioritizing metabolic heat generation over adenosine triphosphate (ATP) production for thermoregulation may lead to the upregulation of AgRP mRNA only when ATP levels are low, resulting in the inhibition of $\alpha$-MSH specifically under such conditions, which could induce pattern transition in response to environmental/seasonal changes [Plotkin et al., 2022; Shimizu et al., 2008] (it should be noted that it has not been demonstrated in wild boars). Additionally, lions are suggested to have different thermoregulation mechanisms from wild boars, such as using BAT/UCP1, similar to the Siberian tigers (*Panthera tigris altaica*) and leopards (*Panthera pardus*) which belong to the same genus [Berg et al., 2006; Gaudry et al., 2017b]. Lions are often associated with habitats where they can avoid extremely cold conditions as adults. In contrast, Siberian tigers must endure such cold environments even in adulthood, making the maintenance of metabolic heat crucial for survival [Carroll and Miquelle, 2006]. Interestingly, Siberian tigers retain the same body patterns into adulthood as they had when they were juveniles. In contrast, while leopards also maintain their juvenile patterns, their habitat differs significantly from that of lions and Siberian tigers, as it is vast [Zeng et al., 2022]. Therefore, survival strategies similar to those of Siberian tigers in extreme cold environments may have been prioritized, potentially explaining why leopard body patterns do not transition from juvenile to adult stages. Given the above considerations, there might be a correlation between pattern transitions and thermoregulation mechanisms or habitat. From a mathematical perspective, the study [Krause et al., 2019] reports cases where spotted patterns can be maintained without being influenced by domain growth. Therefore, even for striped patterns, it might be possible for striped patterns to be maintained without being affected by domain growth if appropriate reaction terms are selected. However, it's important to recognize that this could merely be a potential aspect for studying pattern transition mechanisms, and not necessarily a definitive factor.

On the other hand, in the case of sika deer, pattern transitions are synchronized with habitat changes: metabolism increases in summer and decreases in winter. Under the assumption that the metabolic rates increase/decrease during summer/winter, the system ( 13 ) reproduces this pattern transition. Especially in winter, there are individuals where spots remain on the body surface to some extent, and it is worth considering that this difference is reproduced by the slight difference in metabolic rates, $\kappa\lambda$.

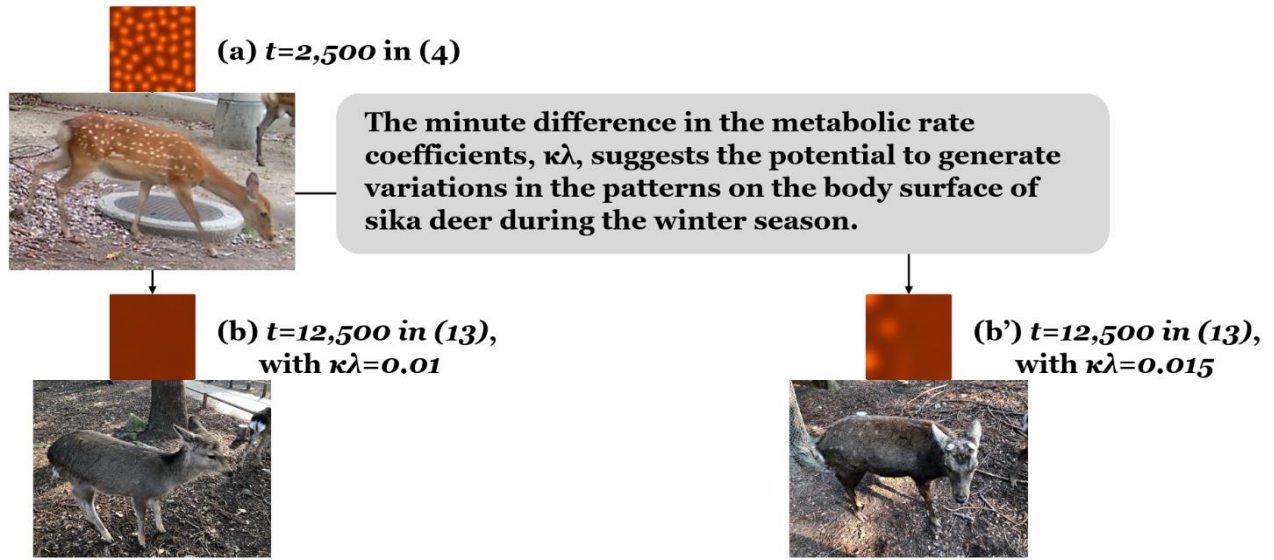

**Figure 3.** The same results of the numerical analysis as those of Fig. 2, with the actual photograph of deer (in September, 2023 (the upper one) and February, 2024 (the bottom ones)). (Figures and Photos in Color)

Moreover, it has been confirmed that elk (*Cervus elaphus nelson*) have lower metabolic rates in specific environments [Parker et al., 1984]. There may be a similar mechanism in sika deer. And this survival strategy of decreasing metabolic rates may correspond to the possibility that the total number of respiratory cycles throughout life remains constant [Escala et al., 2022]. However, it should be noted again that this is not a conclusively proven conclusion.

In an exceptional case differing from the spotted patterns, when changing $\alpha_v$, the coefficient representing the anisotropy in the horizontal direction (*y*-axis direction in this case) of Fig. 1, from 1.0 to 0.7 (it is assumed to diffuse from the dorsal line [Bard, 1981]), the striped patterns formed in the system (4) disperse at $t = 17,000$ (when the domain has grown to approximately 10.2 times its original size) both in the systems (7) and (8) and in the system (12) (see Appendix-C). As illustrated by the unusual case, it's worth noting that patterns may not necessarily disperse more rapidly due to the combined effects of GD and MR. Additionally, as the value of $\alpha_v$ decreases, numerical analysis confirms that striped patterns are less prone to dispersion. Malayan tapirs, like American tapirs, have such striped patterns formed during their juvenile stage that transition as they grow into adults. However, unlike American tapirs, Malayan tapirs have distinct white and black areas that are clearly demarcated in all individuals, as one of notable exceptions [Othmer et al., 2009]. For example, as shown in Fig. 4, if only the value of $\varepsilon$ in a certain domain ($\Omega_W$ in Fig. 4) changes during the growing-domain stages, meaning the production reaction rate of substance $V$ varies based on the equations (1) and (4), a pattern resembling white and black domains will form. Similarly, the patterns on the surface of the calves are not uniformly stripes; rather, those of the head, forelimbs, and hindlimbs are spots. As mentioned in previous studies [Woolley et al., 2021], theoretically, simpler patterns should be formed on smaller domains such as the head and limbs compared to the torso. It might also be possible to induce spotted patterns in these domains, by employing spatial heterogeneity [Woolley et al., 2021], *i.e.*, by altering the value of $\alpha_v$, for instance. However, the exact properties of substance $V$, as well as the reasons for the fluctuations in its production reaction rate and anisotropy, remain uncertain and are subject to theoretical speculation.

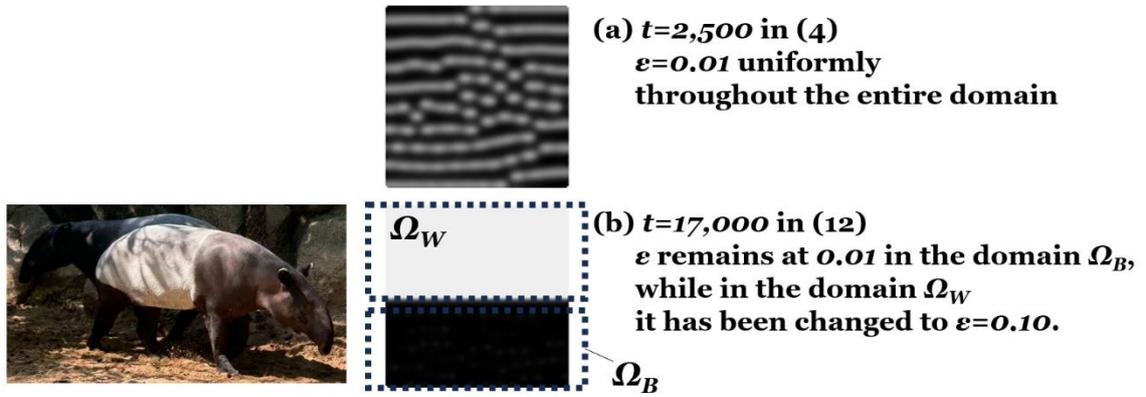

**(a)** $t$=*2,500* in **(4)**
$\varepsilon$=*0.01* uniformly
throughout the entire domain

**(b)** $t$=*17,000* in **(12)**
$\varepsilon$ remains at *0.01* in the domain $\Omega_B$,
while in the domain $\Omega_W$
it has been changed to $\varepsilon$=*0.10*.

$\Omega_W$

$\Omega_B$

**Figure 4.** The numerical analysis results: (**a**) the pattern formed at $t = t_g = 2{,}500$ in the system (4) under the same parameters as those of Fig.1a, (**b**) the pattern formed at $t = 17{,}000$ in the system (12) with $\varepsilon|_{\Omega_W}$ changed from $0.01$ to $0.10$ ($\varepsilon|_{\Omega_B}$ unchanged). The figure of (**a**) resembles the pattern of the Malayan tapir's juvenile, while that of (**b**) resembles that of the adult (in January, 2024). (Photo in Color)

## 6. Author contributions

**Shin Nishihara:** Conceptualization, Methodology, Software, Validation, Formal analysis, Investigation, Resources, Data Curation, Writing - Original Draft, Writing - Review and Editing, and Visualization

**Toru Ohira:** Resources, Writing - Review and Editing, Supervision, Project administration, and Funding acquisition

## 7. Acknowledgements

This work was supported by JSPS Topic-Setting Program to Advance Cutting-Edge Humanities and Social Sciences Research Grant Number JPJS00122674991, and by Ohagi Hospital, Hashimoto, Wakayama, Japan.

## 8. Data and materials availability statements

What supports the findings of this study, such as details of numerical analysis simulations on the relationship between anisotropy and resistance to pattern dispersion, is openly available in arXiv at https://doi.org/10.48550/arXiv.2403.05741.

## Appendix

### A. The Conditions of Turing Instability in the System (4)

Since the equilibrium state $(u^*, v^*)$ in the system (4) satisfies the equation described below:

$$u^* = \frac{\beta + \varepsilon - \delta v^*}{\beta} \,, \tag{A1}$$

$v^*$ is determined by the following equation based on the reaction terms in the system (4):

$$\delta v^3 - (\beta + \varepsilon)v^2 + \beta \delta v - \beta \varepsilon = 0 \,. \tag{A2}$$

Expressing $f_u^*, f_v^*, g_u^*,$ and $g_v^*$ in terms of $v^*$ as shown below with $N \in \mathbb{N}$ ($N = 2$ in the system (4)):

$$f_u^* = \mu(-v^{*N} - \beta) < 0, \qquad f_v^* = \mu(-Nu^*v^{*N-1}) = -N\mu\left(\delta - \frac{\varepsilon}{v^*}\right) < 0, \qquad g_u^* = \mu v^{*N} > 0,$$

$$g_v^* = \mu(Nu^*v^{*N-1} - \delta) = \mu\left((N-1)\delta - \frac{N\varepsilon}{v^*}\right), \tag{A3}$$

$\phi_v f_u^* + \phi_u g_v^* > 0$ in the equation (5) requires $g_v^* > 0$, leading to the condition: $N \geq 2$. While the case of $N = 2$ is discussed in the main body of this paper, the cases of $N = 3,4$, for example, in the system (4) also can form Turing patterns under the parameter spaces: $(\phi_u, \phi_v, \beta, \delta, \varepsilon) \coloneqq (0.21, 0.05, 0.05, 0.12, 0.02)$ and $(0.21, 0.05, 0.03, 0.06, 0.01)$ for $N = 3$ and 4, respectively.

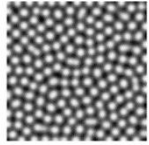     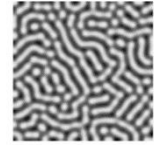

**(a) N=3**                    **(b) N=4**

**Figure A-1.** The numerical analysis results: (**a**) the pattern formed at $t = 1,000$ in the system (4) under the parameters: $(\phi_u, \phi_v, \beta, \delta, \varepsilon, \mu, N) = (0.21, 0.05, 0.05, 0.12, 0.02, 1.0, 3)$, (**b**) that at $t = 1,000$ in the systems (4) under the parameters: $(\phi_u, \phi_v, \beta, \delta, \varepsilon, \mu, N) = (0.21, 0.05, 0.03, 0.06, 0.01, 1.0, 4)$

It should be noted that Turing Instability is never satisfied in the case of $N = 1$, so it is not taken into account in this paper.

### B. Slower Growth and Pattern Transitions in the systems (4), (7), (8) and (12)

For slower growth, the numerical analysis below indicates a phenomenon similar to that shown in Fig. 1, although the time needed for the patterns to disperse is extended.

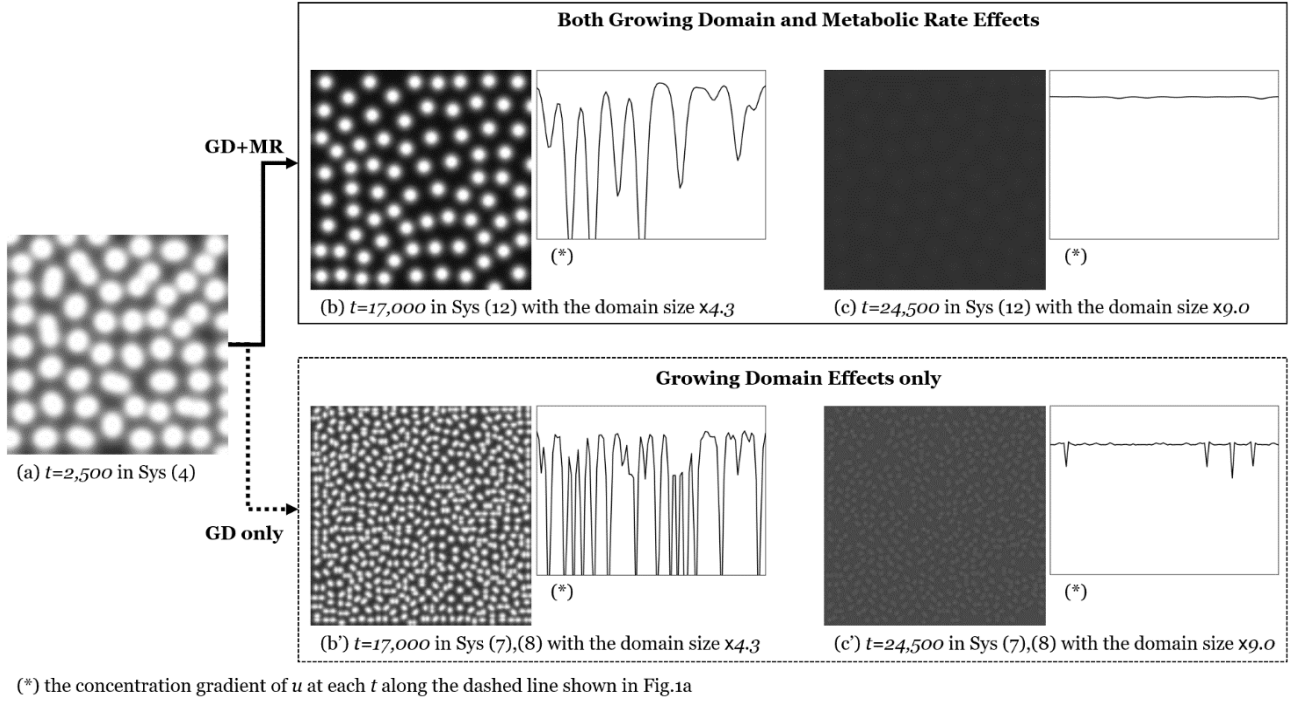

(*) the concentration gradient of $u$ at each $t$ along the dashed line shown in Fig.1a

**Figure B-1.** The numerical analysis results: (**a**) the pattern formed at $t = t_g = 2,500$ in the system (4), (**b**)(**c**) those at $t = 17,000$, and $24,500$ in the system (12), (**b'**)(**c'**) those at $t = 17,000$, and $24,500$ in the systems (7) and (8), under the parameters:

$\left(\phi_u, \phi_v, \beta, \delta, \varepsilon, d_a, \mu\left(\text{the initial value in the system (4)}\right), \alpha_u, \alpha_v, \phi_\tau, \alpha_\tau, \kappa\lambda, \tau_{th}, \gamma\right) =$

$(0.21, 0.05, 0.07, 0.18, 0.01, 2.0 \times 10^{-4}, 0.5, 1.0, 1.0, 0.24, 1.0, 0.2, 0.5, 300)$.

## C. Striped Pattern Transitions in the systems (4), (7), (8) and (12)

In contrast to spotted patterns, the numerical findings below suggest that the time needed for pattern dispersion is nearly the same when both GD and MR effects are present compared to when only GD effects are considered.

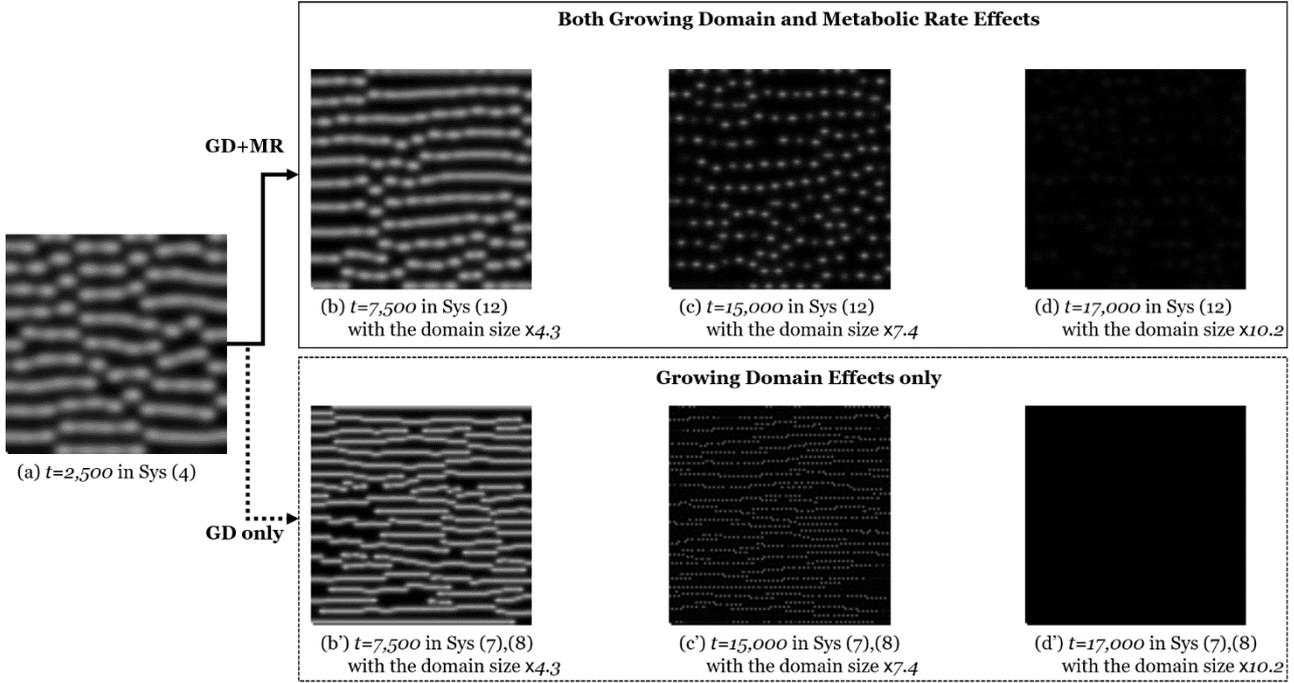

**Figure C-1.** The numerical analysis results: (**a**) the pattern formed at $t = t_g = 2{,}500$ in the system (4), (**b**)(**c**)(**d**) those at $t = 7{,}500$, $15{,}000$, and $17{,}000$ in the system (12), (**b'**)(**c'**)(**d'**) those at $t = 7{,}500$, $15{,}000$, and $17{,}000$ in the systems (7) and (8), under the parameters:
$\left(\phi_u, \phi_v, \beta, \delta, \varepsilon, d_a, \mu(\text{the initial value in the system (4)}), \alpha_u, \alpha_v, \phi_\tau, \alpha_\tau, \kappa\lambda, \tau_{th}, \gamma\right) =$
$(0.21, 0.05, 0.07, 0.18, 0.01, 3.2 \times 10^{-4}, 0.5, 1.0, 0.7, 0.24, 1.0, 0.2, 0.5, 300).$

## D. The Relationship between Anisotropy and Pattern Dispersion

As the value of $\alpha_v$ indicating anisotropy decreases, there is an observed tendency for the time ($t$) taken for patterns to disperse to increase, as demonstrated by the numerical analysis results below:

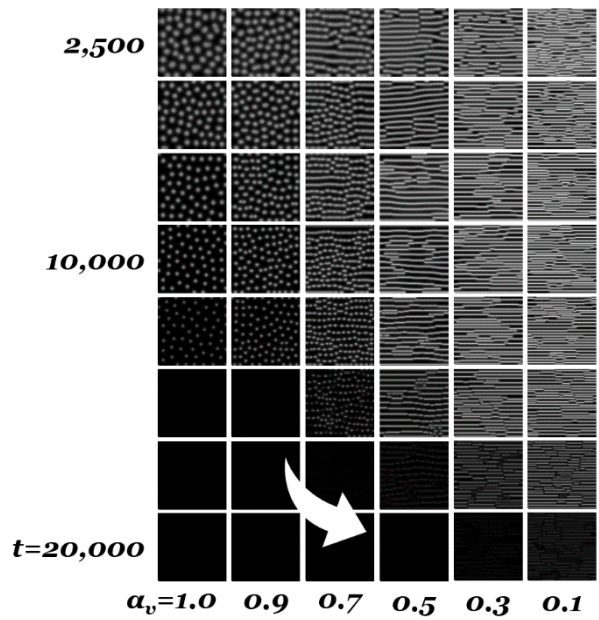

**Figure D-1.** The numerical analysis results: the patterns formed and dispersed at $t = 2,500, 5,000, 7,500, 10,000, 12,500, 15,000, 17,500,$ and $20,000$ in the system $(12)$, under the parameters: $\left(\phi_u, \phi_v, \beta, \delta, \varepsilon, d_a, \mu(\text{the initial value in the system } (4)), \alpha_u, \phi_\tau, \alpha_\tau, \kappa\lambda, \tau_{th}, \gamma\right) = (0.21, 0.05, 0.07, 0.18, 0.01, 3.2 \times 10^{-4}, 0.5, 1.0, 0.24, 1.0, 0.2, 0.5, 300).$